# Precipitation-Redispersion of Cerium Oxide Nanoparticles with Poly(Acrylic Acid) : Towards Stable Dispersions

A. Sehgal[(1,*)], Y. Lalatonne[(1)], J.-F. Berret[(2,*)] and M. Morvan[(1)]

[(1)] : Complex Fluids Laboratory, CNRS - Cranbury Research Center Rhodia Inc., 259 Prospect Plains Road, Cranbury NJ 08512 USA

[(2)] : Matière et Systèmes Complexes, UMR CNRS n° 7057, Université Denis Diderot Paris-VII, 140 rue de Lourmel - 75015 Paris - France

**Abstract** We exploit a precipitation-redispersion mechanism for complexation of short chain polyelectrolytes with cerium oxide nanoparticles to extend their stability ranges. As synthesized, cerium oxide sols at pH 1.4 consist of monodisperse cationic nanocrystalline particles having a hydrodynamic diameter of 10 nm and a molecular weight 400000 g·mol$^{-1}$. We show that short chain uncharged poly(acrylic acid) at low pH when added to a cerium oxide sols leads to macroscopic precipitation. As the pH is increased, the solution spontaneously redisperses into a clear solution of single particles with an anionic poly(acrylic acid) corona. The structure and dynamics of cerium oxide nanosols and their hybrid polymer-inorganic complexes in solution are investigated by static and dynamic light scattering, X-ray scattering, and by chemical analysis. Quantitative analysis of the redispersed sol gives rise to an estimate of 40 - 50 polymer chains per particle for stable suspension. This amount represents 20 % of the mass of the polymer-nanoparticle complexes. This complexation adds utility to the otherwise unstable cerium oxide dispersions by extending the range of stability of the sols in terms of pH, ionic strength and concentration.

*Corresponding authors : jean-francois.berret@ccr.jussieu.fr / Amit.Sehgal@us.rhodia.com

## Introduction

Critical emerging nanomaterials utilize not only the chemical composition but also the size, shape and surface dependant properties of nanoparticles in novel applications with remarkable performance characteristics. Smaller than the wavelength of visible light, these nanoparticles have an important role in a broad range of applications in materials science and in biology [1,2]. In material science, they are used in catalysis or as precursors for ceramics and electro-optic devices. In biology, magnetic nanoparticles provide contrast in magnetic resonance imaging [3-5], and fluorescent quantum dots can be used for biomedical diagnostics [6-8] and cell imaging [9,10]. In order to translate intrinsic properties of nanoparticles to different uses there is a need for a robust means to stabilize nanoparticle dispersions in aqueous media in a variety of processing conditions.

Metal oxides form a class of special interest among inorganic nanoparticles. Some primary examples are oxides of cerium [11-15], iron [4,5,16-19], or zirconium [20-24]. Aqueous dispersions of such metal oxide particles exhibit some common properties: *i)* the particles have an ordered crystalline structure, *ii)* the sols are typically synthesized in extremely acidic (or basic) conditions, *iii)* the particles are stabilized by electrostatics and are extremely sensitive to perturbations in pH, ionic strength and concentration that may dramatically modify their thermodynamic stability [1,11,20,21,25-29]. Destabilization may result from the high surface-to-volume ratio for these particles and from the strong reactivity of the surface chemical sites to physico-chemical changes.

The general limitation for utility of these nanoparticles remains that of their colloidal stability. There are several low-molecular weight molecules that are able to deal efficiently with this issue. These molecules called dispersants, ligands or peptizers serve their role to modify the interactions between nanocolloids through adsorption on the surfaces [20-23,30,31]. The adsorbed molecules result in additional steric or electrostatic barriers for stable sols. For the class of metal oxides above, some of the well-known low-molecular weight molecules are citric acid [21] and poly(acrylic acid) (PAA) [20]. These polyfunctional molecules, three carboxylic acids (COOH) in the case of the citric acid and one COOH per monomer for PAA, form complexes with the surface hydroxyls of the nanocrystals. In many examples however, adding dispersant is not sufficient and the redispersion is insured by mechanical methods such as shear and high-energy ultrasound [30,31]. In this study we address the issue of the adsorption and of the complexation between 10 nm cerium oxide nanoparticles and short PAA chains (molecular weight 2000 g·mol$^{-1}$). We demonstrate that we may extend the range of pH and concentration stability of cerium nanosols considerably by irreversibly adsorbing weak polyelectrolytes on the surface. This process defined as the precipitation-redispersion (P-R) phenomenon and does not require mechanical stimulation.

## II - Experimental

The cerium oxide nanoparticles ($\rho$ = 7.1 g·cm$^{-3}$) suspensions were synthesized as cationic fluorite-like nanocrystals in nitric acid at pH 1.4. The synthetic procedure involves thermohydrolysis of an acidic solution of cerium-IV nitrate salt Ce(NO$_3$)$_4$ at high temperature, resulting in homogeneous precipitation of a cerium oxide nanoparticle "pulp" [32]. The size of the particles was controlled by the excess of hydroxide ions added during the thermohydrolysis. The "pulp" contains relatively monodisperse particles that redisperse spontaneously (peptization) on dilution. The cerium oxide dispersions were first characterized with respect to their ionic strength. For pH 1.4 nanosols, the ionic strength arises almost exclusively from the residual nitrate ions. The ionic strength was found to be 0.044 M, with a Debye screening length $\lambda_D = 1.47$ nm with the usual definition by the DLVO formalism [33].

All cerium oxide nanoparticle batches investigated in this work were thoroughly characterized by light and x-ray scattering experiments. Static and dynamic light scattering were performed on a Brookhaven spectrometer (BI-9000AT autocorrelator, incident wave-length $\lambda$ = 488 nm) for measurements of the Rayleigh ratio $R_\theta(\mathbf{c})$ and of the collective diffusion constant D(**c**) where **c** is the mass fraction of cerium oxide [34]. The Rayleigh ratio measured as a function of the concentration, yielded a weight-average molecular weight $M_w$ (= 400000 g·mole$^{-1}$) of the nanoparticles. The collective diffusion coefficient D(**c**) was measured in the range c = 0.01 – 10 wt. % and the hydrodynamic diameter $D_H$ (= 9.8 ± 0.2 nm) was evaluated from zero concentration extrapolation. For the nanoparticles at pH 1.4, the polydispersity index was estimated by cumulants analysis to be 0.1.

Wide-angle x-ray scattering (WAXS) was utilized to confirm the crystalline structure of the nanocrystals. WAXS was performed at scattering angles between 10 and 65 degrees and revealed the 5 first crystallographic orders of the fluorite structure (of symmetry $Fm\bar{3}m$), which is the structure of cerium oxide [14,35]. The small-angle x-ray scattering (SAXS) experiments were carried out at the Brookhaven National Laboratory on the X21 beam line at the incident wavelength 0.176 nm and sample-to-detector distance 1 meter, to cover the wave-vector range 0.008 – 0.5 Å$^{-1}$. The x-ray scattering data allowed estimation of the radius of gyration $R_G$ (= 3.52 ± 0.02 nm) of the particles. We recall that for monodisperse and homogeneous spheres of radius R, R = $R_H$ (hydrodynamic radius) = 1.29 $R_G$. Here, for the ceria nanoparticles we find $R_H/R_G$ = 1.39, a result that can be explained by the polydispersity in size of the particles. Investigated with high resolution transmission electron microscopy, the nanoceria were found to consist of isotropic agglomerates of 2 - 5 crystallites fused together. Each sub-crystallite have an average size of 2 nm and facetted morphologies, resulting in an overall diameter of 10 nm [11,12]. In this work, it was checked that the structural properties of the cerium particles at the nanometer scale did not affect the precipitation-redispersion process.

As synthesized, the cerium oxide nanoparticles are stabilized merely by electrostatic repulsion and an increase of the pH *or* ionic strength results in aggregation of the particles, and destabilization of the sols [11,28]. As described in Fig. 1, systematic variation of pH by addition of NH$_4$OH over a broad range between 1.4 and 11 resulted in macroscopic phase separation. Though the change in pH apparently reduces the effective surface charge of the particles, it also reduces the range of the electrostatic repulsion due to reduced $\lambda_D$. It is significant to note that the destabilization of the sols occurs well below the point of zero charge of the ceria particles, p.z.c. = 7.9 [11,12]. Based on visual observations, the solution phase behavior may be categorized into 3 different zones.
- Zone 1 - pH = 1.4 to 3 : clear solution
- Zone 2 - pH = 3.5 to 5 : turbid solution that phase separates into a loose precipitate
- Zone 3 - pH = 7 to 11 : rapid phase separation of a dense precipitate

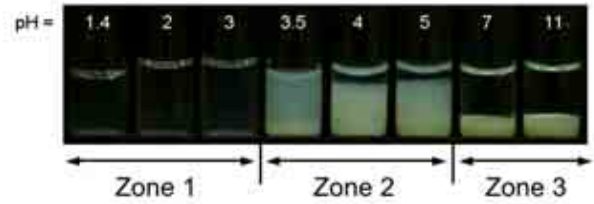

**Figure 1 :** Behavior of the CeO$_2$ sol as function of pH. The initial conditions (first vial on the left) are at pH 1.4 and weight concentration c = 1 wt. %. The suspension pH is adjusted with reagent-grade ammonium hydroxide (NH$_4$OH). Above pH 3.5, a macroscopic phase separation is observed through the apparition of a white precipitate (bottom phase).

This result is reminiscent of the investigations by Nabavi *et al.* [11] where between pH 3 - 5 these authors identified a zone of reversible aggregation and the retention of strongly bound nitrates on the particle surfaces. Moreover, between pH 7 - 10.5, the nitrate layer is lost resulting in rapid and irreversible precipitation. In the following section, we investigate the interactions between nanoceria and PAA. Poly(acrylic acid) with a weight-averaged molecular weight of 2000 g·mol$^{-1}$ and a polydispersity of 1.74 was purchased from Sigma Aldrich. The polymers were dialyzed against deionized water (pH 3) and freeze-dried before use. Titration experiments (data not shown) confirmed that this weak acid has a pKa around 5.5. The suspension pH was adjusted with reagent-grade nitric acid (HNO$_3$) and with ammonium hydroxide (NH$_4$OH).

## III - The precipitation-redispersion phenomenon

The inherent instability of inorganic nanoparticle sols displayed in Fig. 1 may be resolved by complexation with charged ion-containing polymers. To achieve this goal for sol stability, we have developed a two-step process, defined as the precipitation-redispersion process.

**Step #1 - Precipitation** : When cerium oxide sols are mixed with poly(acrylic acid) solutions, both being prepared at the same concentration c and pH 1.4, the solution undergoes an instantaneous and macroscopic precipitation. Mixing of the two initial solutions may be characterized by the ratio $X = V_{CeO_2}/V_{PAA_{2K}}$, where $V_{CeO_2}$ and $V_{PAA_{2K}}$ are the volumes of the cerium and polymer solutions, respectively. The solution becomes turbid on mixing. Sedimentation or centrifugation gave two distinctly separated phases, as shown by the left hand-side vial of Fig. 2a. The bottom phase appears as a yellow and dense precipitate whereas the supernatant is fluid and transparent. From its visual aspect, the two-phases resemble those obtained with bare nanoparticles at high pH (Fig. 1). This precipitation by addition of PAA has been observed for broad ranges in concentrations (c = 0.01 – 10 wt. %) and mixing ratios (X = 0.01 – 100). This process was also investigated using poly(acrylic acid) of different molecular weights, $M_w$ = 2000, 8000 and 30000 g·mol$^{-1}$. In the following, we focus on the 2000 g·mol$^{-1}$ chains (PAA$_{2K}$). Phase separation may result from the multisite adsorptions of the uncharged acrylic acid moieties on the particle surfaces. The exact mechanism for the adsorption (hydrogen bonding or electrostatic) is not known at this stage. Though the PAA is uncharged at pH 1.4 the adsorption was accompanied by a decrease in pH due to the release of excess acid. Complexation or sharing of the polymer chains with several particles results in macroscopic precipitation due to colloidal bridging which is reminiscent of classical associative phase separation [1,36].

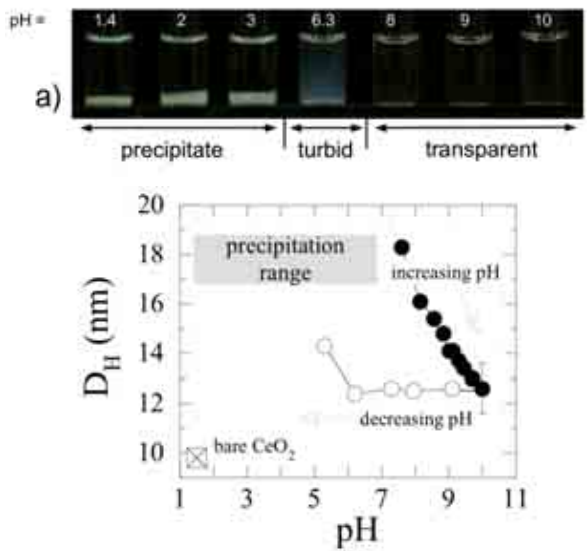

**Figure 2 :** a) Series of CeO$_2$-PAA$_{2K}$ solutions prepared at different pH. The mixing of the polymer and cerium solutions was made at pH 1.4, and the pH was further adjusted with NH$_4$OH. Above pH 7, the precipitate has been redispersed. b) Hydrodynamic diameters measured by dynamic light scattering on CeO$_2$-PAA$_{2K}$ solutions during the Precipitation-Redispersion (close symbols). At pH 10, the PAA$_{2K}$-coated cerium sol is stable and it can be brought back to pH 6 without noticing any change in the dispersion (empty symbols). Below pH 6, the nanoparticles start to aggregate.

**Step #2 - Redispersion :** Starting from the two-phase coexisting state of the precipitate and supernatant as described previously, the pH was then progressively increased by addition of ammonium hydroxide NH$_4$OH. Under steady stirring, the titration by addition of base above pH 7.5 resulted in a complete and rapid (within seconds) redispersion of the precipitate. This evolution of the CeO$_2$-PAA$_{2K}$ solutions at X = 1 and c = 1 wt. % is illustrated in Fig. 2a for pH range from 1.4 and 10. The precipitation-redispersion was followed by static and dynamic light scattering experiments. With increasing pH, there is first a broad region between pH 1.5 and 7 where the precipitate still persists (Fig. 2b). In this range, classical light scattering experiments could not be performed due to multiple scattering. Above pH 7, the solution turns transparent and the autocorrelation function of the scattered light may be measured. In this pH-range, the autocorrelation function was characterized by a slightly polydisperse relaxation mode, associated with an hydrodynamic diameter D$_H$ calculated through the Stokes-Einstein relationship [34].

D$_H$ was found to decrease progressively from 18 nm at pH 7.5 to 12.5 ± 1 nm at pH 10 (Fig. 2b). This latter value is ~ 3 nm larger than the diameter of the bare particles (D$_H$ = 9.8 nm), suggesting that the nanoceria are now coated by the poly(acrylic acid) chains. Doublets, triplets etc…, multiplets of particles would have produced much larger hydrodynamic diameters, or much broader polydispersity. This hypothesis was corroborated by ζ-potential measurements (Zetasizer 3000, Malvern Instrument) which clearly showed a shift from ζ = + 40 mV for the cerium nanosol at pH 1.4 (c = 0.1 wt. %) to ζ = - 45 mV for mixed CeO$_2$-PAA$_{2K}$ solution at X = 1 and pH 10. At the end of the redispersion process, the sign of the electrostatic charge of the particles was reversed. On decreasing the pH from pH 10 by progressive addition of perchloric acid (uncharging the poly(ammonium acrylate)), the CeO$_2$-PAA$_{2K}$ sols remained thermodynamically stable down to pH 6. At still lower pH, a phase separation was again observed as the hydrodynamic diameter and Rayleigh ratio (data not shown) diverge. The hysteresis loop in Fig. 2b illustrates that CeO$_2$-PAA$_{2K}$ sols under the same physico-chemical conditions exist under different colloidal states. In the precipitation-redispersion, PAA$_{2K}$ chains have two distinct functions, one that deals with the redispersion process and the other with the stabilization of the particles. This is explained further through the development below.

## IV – Composition and Structure in the Redispersed State

*Total Organic Carbon Analysis (TOC)* : To further elucidate the P-R phenomenon, the partitioning of the PAA and the composition of the different phases were further investigated by total organic carbon analysis (TOC). A series of CeO$_2$-PAA$_{2K}$ solutions were prepared at pH 1.4 with X = 0.01 to 100. After centrifugation at 15000 rpm, both the supernatant and precipitate were separated and studied by this technique. Fig. 3a shows the residual PAA left in the supernatant *i.e.* uncomplexed during the precipitation as a function of X. The figure emphasizes that for X ≥ 2 all the chains have been complexed. Parallel UV-visible adsorption corroborates the presence of PAA in the upper phase at low mixing ratios, and more importantly that in the whole X-range explored here nano-CeO$_2$ are all incorporated to the dense phase. Estimates of the compositions of the dense phase from TOC analysis show that 80 to 90 % by weight of the precipitate (depending on X) is water whereas the percentages of nanoparticles and polyelectrolytes are in the range 1 – 10 % (Fig. 3b). The precipitate is described as a loose percolated fractal aggregate made of particles and PAA chains. The data in Fig. 3b allows us to calculate the number $n_{ppt}$ of absorbed PAA chains per particle in the precipitate. This number remains constant at $n_{ppt}$ = 140 ± 20 at low values of X and decreases with increasing X (Fig. 3c). This behavior is compared to the nominal number of polymer per particle calculated from X according to :

$$n(X) = \frac{M_w^N}{M_w^P} \frac{1}{X} \qquad \text{Eq. 1}$$

Eq. 1 is valid only when the solutions are mixed at identical concentrations according to the definition above. As anticipated, at large X (X > 2), the experimental and the calculated X's are identical. For the system $CeO_2$-$PAA_{2K}$, this value of X ~ 2 is critical since it is also the mixing ratio below which the redispersion (by raising pH, see previous section) is complete. Incomplete means here that these solutions never become fully transparent and redispersed, even at high pH. This critical X sets the lower limit for redispersion in terms of number of polymer per particle. From the TOC analysis, we conclude that for the $CeO_2$-$PAA_{2K}$ precipitate to be redispersable by pH-increase, nanoceria need to be complexed at least by 140 chains, corresponding to a nominal X of 1.4 (Eq. 1). This number is also the number that the system will adopt spontaneously, even if more polymers are added during mixing. As shown in Fig. 3, the uncomplexed PAA chains remain in the supernatant.

chains (~ 40) on the particle necessary for a stable suspension is therefore less than the added PAA.

*Static Light Scattering* : Static light scattering has been utilized to determine the molecular weight of the $PAA_{2K}$-coated nanoparticles and hence to derive the number $n_{ads}$ of adsorbed polymers per particle. To do so, 200 mL of a $CeO_2$-$PAA_{2K}$ solution have been prepared at X = 1, c = 0.2 wt. % and pH 10. The solution was then ultra-filtrated in a cell equipped with a 3000 g·mol$^{-1}$ pore size filter (Pall Life Sciences) in order to remove the free counterions and residual uncomplexed PAA. It was checked at the end of the ultra-filtration that the redispersed sol was characterized by 14 ± 1 nm hydrodynamic diameter monodisperse particles. This value is in agreement with the findings of Fig. 2.

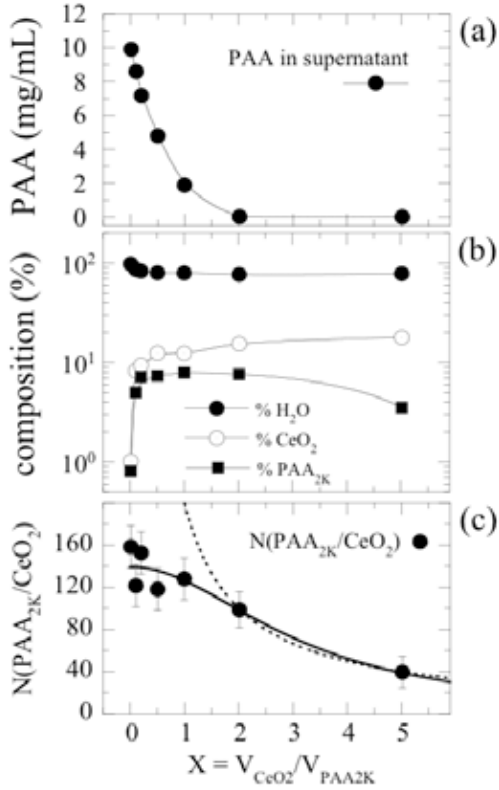

**Figure 3** : a) Residual poly(acrylic acid) in the supernatant as function of X for a series of $CeO_2$-$PAA_{2K}$ mixed solutions prepared at c = 1 wt. % and pH 1.4. Prior to the measurements by Total Organic Content analysis (TOC), the suspensions were centrifuged at 15000 rpm. b) Composition by weight of the precipitates as function of X for the same solutions as in Fig. 3a. c) Number $n_{ppt}$ of adsorbed PAA chains per particle in the precipitate. $n_{ppt}$ is calculated from the data in Figs. 3a and 3b.

To exclude the uncomplexed PAA the redispersed sol was dialyzed through a 10000 g·mol$^{-1}$ membrane and TOC analysis was extended to the supernatant. The analysis surprisingly revealed that the dialyzed $CeO_2$ – $PAA_{2K}$ dispersion contained only 39 PAA chains per particle i.e. 20% by mass of the complex. Apparently some PAA may be released during the redispersion process. Though 140 $PAA_{2K}$ chains are necessary for redispersion to occur the number of absorbed $PAA_{2K}$

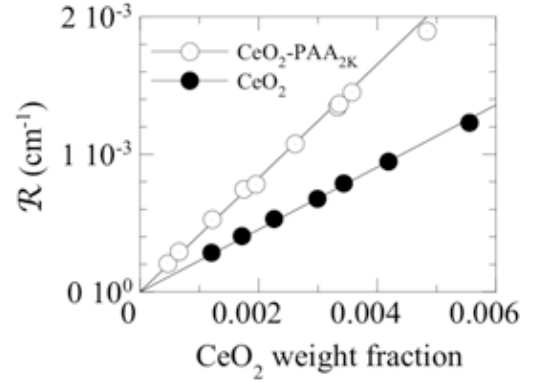

**Figure 4 :** Rayleigh ratios measured by static light scattering for bare $CeO_2$ particles at pH 1.4 and for $CeO_2$-$PAA_{2K}$ solutions redispersed at pH 10. The concentration in abscise is that of the bare particles. These data allow us to derive the number of adsorbed polymers per particle, $n_{ads}$ = 49 ± 5 (from Eq. 3).

Fig. 4 displays the Rayleigh ratios $\mathcal{R}_\theta(c)$ of the PAA-coated particles as function of the $CeO_2$ concentration in the range 0 – 0.6 wt. %. This scattered intensity is compared to that of the bare nanoparticles. In the two cases, $\mathcal{R}_\theta(c)$ varies linearly with c according to [34,37] :

$$\mathcal{R}_\theta(c) = K\, c\, M_w^{app} \qquad \text{Eq. 2}$$

where $M_w^{app}$ is the weight-average apparent molecular weight of the scattering entities and $K = 4\pi^2 n^2 (dn/dc)^2 / N_A \lambda^4$ is the scattering contrast ($N_A$ is the Avogadro number and dn/dc is the refractive index increment, measured on a Chromatix KMX-16 differential refractometer). The ratio between the two straight lines in Fig. 4 is 1.83. This ratio can be calculated as function the characteristics of the two systems, yielding:

$$\frac{\mathcal{R}_\theta^{CeO_2\text{-}PAA_{2K}}(c)}{\mathcal{R}_\theta^{CeO_2}(c)} = \frac{K_{CeO_2\text{-}PAA_{2K}}}{K_{CeO_2}} \left(1 + n_{ads} \frac{M_w^P}{M_w^N}\right)^2$$
Eq. 3

Here $K_{CeO_2-PAA_{2K}}$ (= $0.99 \cdot 10^{-6}$ mol·cm$^{-2}$·g$^{-2}$) and $K_{CeO_2}$ (= $1.21 \cdot 10^{-6}$ mol·cm$^{-2}$·g$^{-2}$) are the contrasts for the coated and bare particles respectively. From the comparison between the data of Fig. 4 and Eq. 3, we find $n_{ads} = 49 \pm 5$.

*Small-Angle X-ray Scattering* : Similarly, SAXS was performed on both bare and PAA$_{2K}$-coated nanoparticles in order to compare the two systems at the nanometer scale. SAXS intensities have been collected for dilution series with concentrations comprised between 0.1 and 5 wt. %. Fig. 5 exhibits the intensities of dilute solutions (c < 0.5 wt. % and X = 1), that is for solutions where no structure factor is apparent. At low concentration, the scattered intensity is proportional to the concentration and the q-dependence of the intensity reflects the form factor of the aggregates. The form factors for the bare and coated CeO$_2$-nanoparticles have been shifted vertically so as to superimpose the scattering cross-sections at high wave-vectors (Fig. 5). This result indicates that on a local scale (*i.e.* for distances below 5 nm) both particles have the same structure. A slight deviation shows up below 0.03 Å$^{-1}$ that is ascribed to the PAA corona surrounding the particles. This extra contribution is also illustrated in the inset of Fig. 5 which shows the intensities in a Guinier representation. Here, the logarithm of the intensity decreases linearly with q$^2$ and from the straight lines [37], we deduced a radius of gyration $R_G = 3.52 \pm 0.02$ nm for the bare nanoparticles and $R_G = 4.8 \pm 0.2$ nm for the PAA$_{2K}$ coated particles.

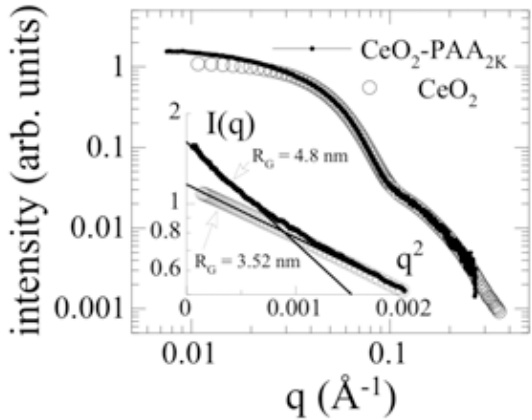

**Figure 5 :** X-ray scattering intensity for bare and PAA$_{2K}$-coated nanoparticles in double logarithmic scale. The concentrations are in the two cases c = 0.5 wt. %. At such low concentrations, the intensities represent the form factors of the particles. The deviation below 0.03 Å$^{-1}$ is due to the PAA corona surrounding the particles. *Inset* : Guinier representation of the intensity for the same samples (I(q) *versus* q$^2$). From the straight lines, the gyration radius for the bare and for the PAA$_{2K}$-coated nanoceria can be calculated, $R_G$ = 3.52 nm and 4.8 nm, respectively.

As $q \rightarrow 0$, the intensity of the coated particles is 1.4 times that of the bare nanoceria. This value can be used to estimate the number of adsorbed polymers. We assume that the coated nanoparticles are of core-shell type, each compartment having an electronic densities $\rho_{CeO_2}$ and $\rho_{PAA_{2K}}$. The excess of intensity as $q \rightarrow 0$ due to the PAA shell with respect to that of the bare particles reads [37,38]:

$$\left.\frac{I_{CeO_2-PAA_{2K}}}{I_{CeO_2}}\right|_{q \rightarrow 0} = \left(1 + n_{ads} \frac{\rho_{PAA_{2K}} - \rho_S}{\rho_{CeO_2} - \rho_S} \frac{V_{PAA_{2K}}}{V_{CeO_2}}\right)^2$$

Eq. 4

where $\rho_S$ are the electronic scattering density of the solvent, $V_{PAA_{2K}}$ is the molar volume of poly(ammonium acrylate) polymer and $V_{CeO_2}$ is that of the bare particle. Using for the densities $\rho_{PAA_{2K}}$ = 0.88 eÅ$^{-3}$, $\rho_{CeO_2}$ = 1.84 eÅ$^{-3}$, $\rho_S$ = 0.33 eÅ$^{-3}$ [39-41], and for the volume of an acrylate monomer $v_0$ = 54.8 Å$^3$ [42], ($V_{CeO_2}$ is calculated from the molecular weight), one ends up with a number of adsorbed polymers $n_{ads} = 34 \pm 6$. Both light and x-ray scattering experiments are in good agreement with each other and with the TOC results. These results indicates that while a critical number of at least 140 polymers is necessary for the redispersion to be realized, only one-third of the PAA's are useful at the stabilization of the particles. We conclude then that during the redispersion, the majority of the polymers are released into the solvent phase. The final amount of polymer per particle is this 40 – 50, which corresponds to 1/5 of the total weight of the coated particles and to a surface coverage of 1 mg·m$^{-2}$.

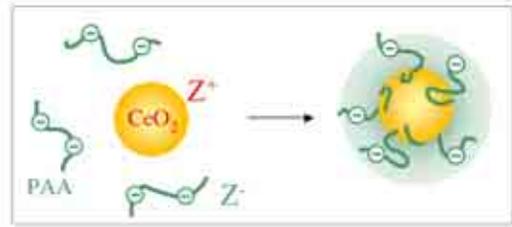

**Figure 6 :** Schematical representation of PAA$_{2K}$-coated cerium nanoparticles obtained through the precipitation-redispersion process.

In order to get a better insight about the conformation of the adsorbed polyelectrolytes, we finally discuss two extremely different configurations and compare them to the present results. In a first case, we assume that all the acrylic acid monomers are adsorbed or in the close vicinity of the cerium oxide surfaces. With an average of 50 chains per particle and a monomeric molar volume $v_0$ = 54.8 Å$^3$ [42], a rapid estimate shows that the radius of the polymer-nanoparticle would then be increased by 0.4 nm. In a second case, we assume that the chains are attached by a few monomers and that the remaining part of the chain is highly stretched towards the surrounding solvent. Counting 0.15 nm per monomer, we end up with a diameter for the whole aggregate of 18 nm. The experimental hydrodynamic diameter $D_H$ = 13 – 14 nm is intermediate and therefore suggests an intermediate conformation for the PAA's. One part of the monomers is adsorbed on the nanoparticles and the remaining part is building a solvated polyelectrolyte brush. Titration of the PAA$_{2K}$-coated nanoparticles from the redispersed state have shown in addition that ~ 1/3 of the total monomers can be deionized by progressive addition of perchloric acid. This value of 1/3 sets the brush thickness to 1.5 nm, again in agreement with the experimental data. The conformation of the adsorbed polyelectrolyte chains is illustrated schematically in Fig. 6.

## V - Conclusion

We have studied the precipitation-redispersion process between cerium nanoparticles and short poly(acrylic acid) chains. By this process, we are able to produce single nanoparticles irreversibly coated with PAA chains (Fig. 6). With a ~ 2 nm polyelectrolyte brush surrounding the particles, the cerium sols are stable over a broad range of pH and concentration. As reported by several authors [1,20], the present results confirmed that the interparticle interactions in the redisperse state can be considered as electrosteric. The highest $CeO_2$–$PAA_{2K}$ concentration achieved so far by ultra-filtration is 50 wt. %, i.e. 500 g·l$^{-1}$, in a stable and transparent solution containing by weight 4/5 of particles and 1/5 of polymers. Results about the ionic strength, concentration, and polymer molecular weights will be described in details in our forthcoming paper. We suggest finally that the precipitation-redispersion process described here could be easily applied to other nanoparticle systems.

**Acknowledgements** : We thank Luzia Novaki, Kenneth Wong (Fisico-Quimica, Rhodia Brazil Ltda.) and Lin Yang (Brookhaven National Laboratory) for their support in the SAXS experiments. This research was carried out in part at the National Synchrotron Light Source, Brookhaven National Laboratory, which is supported by the U.S. Department of Energy, Division of Materials Sciences and Division of Chemical Sciences, under Contract No. DE-AC02-98CH10886. We would like to acknowledge the expertise of the Rhodia rare earths team that synthesized the bare ceria nano sols. AS would like to thank Mitchell Berman for the TOC measurements.